\documentclass[pra,aps,twocolumn]{revtex4}
\usepackage{epsf}
\usepackage{amsmath}
\usepackage{amssymb}

\def \mbf#1{{\mbox{\boldmath$#1$}}} % bold greek

\begin{document}

\title{Optical-state truncation and teleportation
of qudits by\\ conditional eight-port interferometry}

\author{Adam Miranowicz}

\address{Nonlinear Optics Division, Physics Institute, Adam
 Mickiewicz University, 61-614 Pozna\'n, Poland,}

\address{SORST Research Team for Interacting Career Electronics,
Graduate School for Engineering Science, Osaka University, Osaka
560-8531, Japan}

\begin{abstract}
\centerline{published in {\em J. Opt. B: Quantum Semiclass. Opt.}
{\bf 7}, 142-150  (2005)}

\vspace{5mm} \noindent Quantum scissors device of Pegg {\em et al}
(1998 {\em Phys. Rev. Lett.} {\bf 81} 1604) enables truncation of
the Fock-state expansion of an input optical field to qubit and
qutrit (three-dimensional) states only. Here, a generalized
scissors device is proposed using an eight-port optical
interferometer. Upon post-selection based on photon counting
results, the interferometer implements generation and
teleportation of qudit ($d$-dimensional) states by truncation of
an input field at the $(d-1)$th term of its Fock-state expansion
up to $d=6$. Examples of selective truncations, which can be
interpreted as a Fock-state filtering and hole burning in the Fock
space of an input optical field, are discussed. Deterioration of
the truncation due to imperfect photon counting is discussed
including inefficiency, dark counts and realistic photon-number
resolutions of photodetectors.
\\


\end{abstract}

% \pacs{03.65.Ud, 42.50.Dv}

\maketitle

\pagenumbering{arabic}

\section{Introduction} %1

Quantum engineering of nonclassical light has attracted remarkable
interest in the last decade \cite{JMO97}. This interest has
further been stimulated by a recent theoretical demonstration of
Knill {\em et al} \cite{Knill01} that linear optical systems
enable efficient quantum computation. Such systems are
experimentally realizable with present-day technology
\cite{OBrien03,Yamamoto03}, since they are based only on beam
splitters (BSs) and phase shifters (PSs) together with
photodetectors and single-photon sources.

In this paper, linear systems are studied for the {\em
optical-state truncation}, which refers to truncation of the
Fock-state expansion of an input optical state
\begin{eqnarray}
|\psi \rangle =\sum_{n=0}^{\infty }\gamma_{n}|n\rangle, \label{N01}
\end{eqnarray}
with unknown superposition coefficients $\gamma_{n}$, into the
following finite superposition of $d$ states:
\begin{eqnarray}
|\phi^{(d)}_{\rm trun} \rangle ={\cal N}
\sum_{n=0}^{d-1}\gamma_{n}|n\rangle, \label{N02}
\end{eqnarray}
which is called the optical {\em qudit} state ($d$-dimensional
generalized qubit). Here ${\cal N}=(\sum_{n=0}^{d-1}
|\gamma_{n}|^2)^{-1/2}$ is the renormalization constant.  In the
following, the similarity sign will be used instead of writing
explicitly ${\cal N}$. The input state $|\psi \rangle$ can be a
coherent state $|\alpha \rangle$, or any other infinite or finite
dimensional state. Systems for the optical-state truncation are
referred to as the {\em quantum scissors devices} (QSDs).

The first and simplest QSD was proposed by Pegg {\em et al}
\cite{Pegg98,Barnett99}, then analyzed theoretically in various
contexts by others
\cite{Villas99,Koniorczyk00,Paris00,Gui00,Villas01,Ozdemir01,Ozdemir02,Miranowicz04},
and experimentally realized by Babichev {\em et al}
\cite{Babichev03} and Resch {\em et al} \cite{Resch02}. This
device, schematically depicted in figure 1, is composed of linear
optical elements (two beam splitters BS1 and BS2) and
photodetectors D2 and D4 (label 4 is used for consistency with the
other schemes that are discussed in the following). If a
single-photon Fock state $|1\rangle$ is in one of the input modes
$\hat{a}_1$ or $\hat{a}_2$ and the vacuum state is in the other
input, while the measurement has resulted in a single count in one
of the detectors and no count in the other, then the state $|\psi
\rangle$, given by (\ref{N01}), in the input mode $\hat{a}_4$ can
be truncated to the qubit state
\begin{eqnarray}
|\phi^{(2)}_{\rm trun} \rangle  \sim \gamma _{0}|0\rangle +\gamma
_{1}|1\rangle \label{N03}
\end{eqnarray}
in the output mode $\hat{b}_1$, as a special case of (\ref{N02}).
Detailed analysis of the conditions to obtain the target state
(\ref{N03}) as a function of the BS parameters, input states and
measurement outcomes will be reviewed in section 3. The described
optical state truncation based on conditional measurements is
referred to as the {\em projection synthesis} \cite{Pegg98}, which
is a powerful method applied also for other purposes
\cite{Barnett96,Phillips98,Baseia97,Serra00,Pregnell02}. By
analyzing figure 1, it is easy to observe that light from the
input mode $\hat{a}_4$ cannot reach the output mode $\hat{b}_1$.
So, one can conclude that the truncation is achieved via the
quantum teleportation (\cite{Bennett93}, for a review see
\cite{Miranowicz02}), though not of the entire input state but
only of the first two terms of its Fock-state expansion
\cite{Pegg98,Villas99,Koniorczyk00,Lee00,Babichev03}. Thus, in a
special case, if the input field is already prepared in a qubit
state, then the scissors become a conventional teleporting device.
However, contrary to the unconditional teleportation scheme of
Bennett {\em et al} \cite{Bennett93}, the teleportation (and
truncation) via the Pegg-Phillips-Barnett QSD is successful only
in the cases when the two detectors count one photon in total.

The concept of optical-state truncation is by no means limited to
the discussed truncation of the number-state expansion of a given
state. For instance, by considering truncation of a coherent
state, defined by the action of the displacement operator on the
vacuum state, one can truncate the displacement operator and then
apply it to the vacuum state. Such a truncated state is
essentially different (for $d>2$) from that given by (\ref{N02})
\cite{Buzek92,Miranowicz94,Miranowicz01}. Nevertheless, it is
physically realizable, e.g., in a pumped ring cavity with a Kerr
nonlinear medium as was demonstrated by Leo\'nski {\em et al}
\cite{Leonski94,Leonski97,Miranowicz96,Leonski01}. The QSD schemes
can also be generalized for the truncation of two-mode
\cite{Leonski04} or multi-mode fields.

The original Pegg-Phillips-Barnett QSD enables the truncation of
an input state only to qubit and qutrit ($3$-dimensional qudit)
states as was shown by Koniorczyk {\em et al} \cite{Koniorczyk00}.
Here, we discuss a generalization of the QSD for truncation and
teleportation of $d=2,...,6$ dimensional qudits and suggest a way
to extend the scheme for an arbitrary $d$. Our approach is
essentially different from the other qudit truncation schemes
\cite{Villas01,Koniorczyk00,Leonski97,Miranowicz96} and we believe
that it is easier to be experimentally realized. The proposed
scheme is based on an eight-port optical interferometer shown in
figure 2. The setup resembles a well-known multiport
interferometer of Zeilinger {\em et al}
\cite{Zeilinger93,Zeilinger94}, which has been theoretically
analyzed \cite{Reck94,Jex95,Zukowski97,Campos00,Zukowski00} and
experimentally applied \cite{Mattle95,Reck96,Mohseni04} for
various purposes but, to our knowledge, has not yet been used for
optical-state truncation. An important difference between the
standard multiport and that proposed here lies in the elimination
of the apex BS of the triangle. This elimination is important for
the processes of truncation and teleportation.

The paper is organized as follows. In section 2, the generalized
quantum scissors device is proposed including a description of the
setup (2.1), a short review of multiport unitary transformations
(2.2), and explanation of the projection synthesis (2.3), which
enables the qudit state truncation. Reductions of the generalized
QSD to the Pegg-Phillips-Barnett QSD are demonstrated in section
3. Detailed analyses of the generalized QSD for the truncation up
to three, four and five photon-number states are given in sections
4, 6, and 7, respectively. Selective truncations, which can be
interpreted as a Fock-state filtering \cite{Dariano00} or a hole
burning in Fock space \cite{Escher04,Gerry02}, are discussed in
section 5. How imperfect photon counting deteriorates the
truncation processes is discussed in section 8 by including
realistic photon-number resolutions, inefficiency, and dark counts
of photodetectors. Final conclusions with a discussion of open
problems including a generalization of the scheme for an arbitrary
qudit state truncations are presented section 9.

%---------------------------------------------------------------------------
\begin{figure}% scheme 1
\vspace*{5mm} \centerline{\epsfxsize=8cm\epsfbox{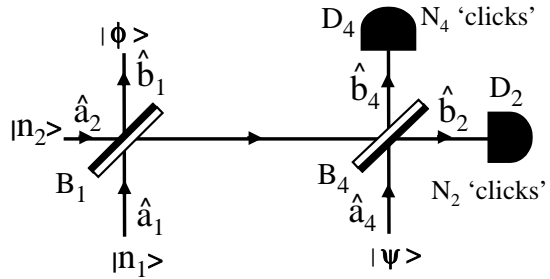}}

\vspace*{-8mm} \caption{Six-port quantum scissors device of Pegg,
Phillips and Barnett. Key: $|\psi\rangle$ -- input state which, in
particular, can be a coherent state $|\alpha\rangle$; $|n_j\rangle$
-- input Fock states; $|\phi\rangle$ -- output qubit or qutrit
state; $D_j$ -- photon counters; $B_j$ -- beam splitters;
$\hat{a}_j$ and $\hat{b}_j$ -- input and output annihilation
operators, respectively.}
\end{figure}

\section{Generalized QSD} %2

\subsection*{\em 2.1 The setup} %2.1

We analyze a generalized quantum scissors device (GQSD) based on
an eight-port optical interferometer, also referred to as a
multiport mixer or multiport beam splitter, which is assembled in
a pyramid-like  configuration of ordinary beam splitters (BSs) and
phase shifters (PSs) as shown in figure 2. The most general
four-port beam-splitter scattering matrix reads as (see, e.g.,
\cite{Campos89,Leonhardt97})
\begin{equation}
{\bf B'}=e^{i\theta_{0}}\left[
\begin{array}{cc}
t \exp(i\theta_t)   & r \exp(i\theta_r)   \\
-r \exp(-i\theta_r)   & t \exp(-i\theta_t)
\end{array}
\right] \label{N04}
\end{equation}
where $T=t^2$ describes the transmittance, and $R=r^2=1-T$ is the
reflectance of the BS. The associated phase factors $\theta_t$ and
$\theta_r$ can be realized by the external phase shifters,
described by ${\bf P}_{\pm }={\rm diag}[\exp (i\theta_t \pm
i\theta_r ),1]$, which are placed in front of and behind the beam
splitter described by real scattering matrix
\begin{equation}
{\bf B}=\left[
\begin{array}{cc}
t  & r  \\
-r  & t
\end{array}
\right] \label{N05}
\end{equation}
as comes from the decomposition
\begin{equation}
{\bf B'}=\exp(i\theta _{0}'){\bf P}_{+}{\bf B}{\bf P}_{-}
\label{N06}
\end{equation}
where $\theta _{0}'=\theta _{0}-\theta_t$. Without loss of
generality the global phase factors $\exp(i\theta _{0}')$ and
$\exp(i\theta _{0})$ can be omitted. Note that (\ref{N05})
describes an asymmetric BS, as marked in all figures by bars with
distinct surfaces. We use the same convention as in
\cite{Ralph02,Skaar04} that beams reflected from the white surface
are $\pi$ phase shifted so, according to (\ref{N05}), the
reflection from the black surface and transmissions from any side
are without phase shift.

The total scattering matrix ${\bf S}$ of the GQSD, shown in figure
2, can be given by
\begin{equation}
{\bf S}={\bf P}_6 {\bf B}_5 {\bf P}_5 {\bf B}_4 {\bf P}_4 {\bf B}_3
{\bf P}_3 {\bf B}_2 {\bf P}_2 {\bf B}_1 {\bf P}_1 \label{N07}
\end{equation}
which is the sequence of two-mode `real' beam splitters described
explicitly by
\begin{eqnarray}
{\bf B}_{1}&=&\left[
\begin{array}{cccc}
t_{1} & r_{1} & 0 & 0 \\
-r_{1} & t_{1} & 0 & 0 \\
0 & 0 & 1 & 0 \\
0 & 0 & 0 & 1
\end{array}
\right], \quad
 {\bf B}_{2}=\left[
\begin{array}{cccc}
t_{2} & 0 & r_{2} & 0 \\
0 & 1 & 0 & 0 \\
-r_{2} & 0 & t_{2} & 0 \\
0 & 0 & 0 & 1
\end{array}
\right], \nonumber \\
{\bf B}_{3}&=&\left[
\begin{array}{cccc}
1 & 0 & 0 & 0 \\
0 & t_{3} & r_{3} & 0 \\
0 & -r_{3} & t_{3} & 0 \\
0 & 0 & 0 & 1
\end{array}
\right], \quad
 {\bf B}_{4}=\left[
\begin{array}{cccc}
1 & 0 & 0 & 0 \\
0 & t_{4} & 0 & r_{4} \\
0 & 0 & 1 & 0 \\
0 & -r_{4} & 0 & t_{4}
\end{array}
\right], \nonumber \\
&&\hspace{12mm}
 {\bf B}_{5}=\left[
\begin{array}{cccc}
1 & 0 & 0 & 0 \\
0 & 1 & 0 & 0 \\
0 & 0 & t_{5} & r_{5} \\
0 & 0 & -r_{5} & t_{5}
\end{array}
\right] \label{N08}
\end{eqnarray}
and phase shifters represented by
\begin{eqnarray}
{\bf P}_{k}&=& {\rm diag}[\exp (i\xi_k),1,1,1] \quad {\rm for}
\quad k=1,2,6,
\nonumber \\
{\bf P}_{k}&=& {\rm diag}[1,\exp (i\xi_k),1,1] \quad {\rm for}
\quad k=3,4,
\nonumber \\
{\bf P}_{5}&=& {\rm diag}[1,1,\exp (i\xi_5),1]. \label{N09}
\end{eqnarray}
Similarly, the diagonal matrix
\begin{equation}
{\bf M}_{k}={\rm diag}[\exp (i\zeta \delta _{1k}),\exp (i\zeta
\delta _{2k}),\exp (i\zeta \delta _{3k}),\exp (i\zeta \delta
_{4k})] \label{N10}
\end{equation}
describes the $k$th mode reflection phase shift caused by the
mirror (see figure 2), where $\delta _{jk}$ is Kronecker delta. We
have not written explicitly ${\bf M}_k$ in the sequence
(\ref{N07}), since the reflection phase shifts $\zeta$ for modes
1, 2 and 3 can be incorporated in $\xi_1$, $\xi_3$ and $\xi_5$,
respectively. Besides the reflection phase shift in mode 4 does
not affect photodetection in D4, and thus can be neglected.

The described setup resembles a well-known multiport interferometer
in triangle configuration of the beam splitters as studied in
various contexts, since its theoretical proposal and experimental
realizations by Zeilinger {\em et al}
\cite{Zeilinger93,Zeilinger94,Reck94,Mattle95}. However, an
important difference between the standard multiport interferometer
and that analyzed here is the elimination of the apex BS of the
triangle, which is usually placed at the crossing of beams 1 and 4
in figure 2. In particular, the setup for the Pegg-Phillips-Barnett
QSD resembles the Zeilinger six-port interferometer with one BS
removed. This elimination is essential for the processes of
truncation and teleportation of qudits as will be shown in the
following.

%---------------------------------------------------------------------------
\begin{figure}% scheme 2
\centerline{\epsfxsize=10cm\epsfbox{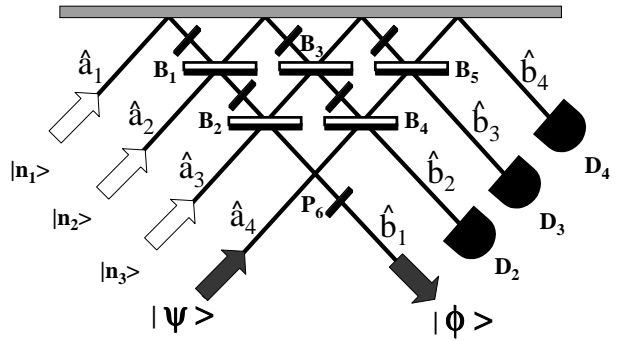}} \vspace*{-1cm}
\caption{Generalized eight-port quantum scissors device. Notation
is the same as in figure 1 but $|\phi\rangle$ denotes the output
qudit state; phase shifters $P_6$ and $P_j$, in front of the beam
splitter $B_j$, are shown by small black bars, while the mirror is
depicted by large bar.}
\end{figure}

\subsection*{\em 2.2 Multiport unitary transformation} %2.2

The annihilation operators $\hat{a}_i$ at the $N$ inputs to
multiport linear interferometer are related to the annihilation
operators $\hat{b}_i$ at the $N$ outputs as follows (see, e.g.,
\cite{Skaar04,Leonhardt97}):
\begin{eqnarray}
\hat{b}_i = \hat{U}^{\dag} \hat{a}_{i}
\hat{U}=\sum_{j=1}^{N}S_{ij}\hat{a}_{j} \label{N11}
\end{eqnarray}
where $S_{ij}$ are the elements of the unitary scattering matrix
${\bf S}$, and $\hat{U}$ is the unitary operator describing the
evolution of the $N$-mode input state, say $|\Psi\rangle$, into the
$N$-mode output state, say $|\Phi\rangle$:
\begin{equation}
|\Phi\rangle=\hat{U}|\Psi\rangle. \label{N12}
\end{equation}
By introducing the column vectors ${\bf \hat{a}}\equiv
[\hat{a}_1;\hat{a}_2; \dots;\hat{a}_N]$ and ${\bf \hat{b}}\equiv
[\hat{b}_1;\hat{b}_2;\dots;\hat{b}_N]$, the set of equations
(\ref{N11}) can compactly be rewritten as
\begin{equation}
{\bf \hat{b}} = \hat{U}^{\dag }\hat{\mathbf{a}}\hat{U} =
\mathbf{S}\,\mathbf{\hat{a}} \label{N13}
\end{equation}
from which follow the inverse relations for the creation operators
\begin{eqnarray}
\mathbf{\hat{a}}^{\dag }=\hat{U}\mathbf{\hat{b}}^{\dag
}\hat{U}^{\dag }= \mathbf{S}^{T}\mathbf{\hat{b}} ^{\dag }.
\label{N14}
\end{eqnarray}
Then, one can observe that
\begin{equation}
\hat{U}\hat{a}_{i}^{\dag }\hat{U}^{\dag
}=\hat{U}(\sum_{j}S_{ji}\hat{b} _{j}^{\dag })\hat{U}^{\dag
}=\sum_{j}S_{ji}\hat{U}\hat{b} _{j}^{\dag }\hat{U}^{\dag
}=\sum_{j}S_{ji}\hat{a} _{j}^{\dag }. \label{N15}
\end{equation}
By applying (\ref{N15}) and noting that neither BSs nor PSs change
the vacuum state, $\hat{U}|{\bf 0}\rangle=|{\bf 0}\rangle$, one can
calculate the output state $|\Phi\rangle$ of the multiport
interferometer described by the scattering matrix ${\bf S}$ for the
input Fock states $|\Psi\rangle=|n_1,\dots,n_N\rangle\equiv |{\bf
n}\rangle$ as follows (for detailed examples see
\cite{Skaar04,Leonhardt97}):
\begin{eqnarray}
\hat{U}|{\bf n}\rangle
&=&\hat{U}\prod_{i=1}^{N}\frac{(\hat{a}_{i}^{\dag
})^{n_{i}}}{\sqrt{ n_{i}!}}|{\bf 0}\rangle
\nonumber \\
&=&\prod_{i=1}^{N}\frac{1}{\sqrt{n_{i}!}}(\hat{U}\hat{a}_{i}^{\dag
}\hat{U}^{\dag })^{n_{i}}|{\bf 0}\rangle  \label{N16} \\
&=&\prod_{i=1}^{N}\frac{1}{\sqrt{n_{i}!}}\left(
\sum_{j=1}^{N}S_{ji}\hat{a}_{j}^{\dag }\right) ^{n_{i}}|{\bf 0}\rangle  \nonumber \\
\nonumber \\
&=&\frac{1}{\sqrt{n_{1}!\cdots n_{N}!}}\sum_{{\bf j }
=1}^{N}\prod_{l=1}^{ {M} }S_{j_{l}x_{l}}\hat{a}_{j_{l}}^{\dag
}|{\bf 0}\rangle \nonumber
\end{eqnarray}
where ${M} =\sum_{i}n_{i}$ is the total number of photons;
$\{x_{j}\}\equiv ( \underbrace{1,...,1}_{n_{1}},
\underbrace{2,...,2}_{_{n_{2}}}, ...,\underbrace{
N,...,N}_{_{n_{N}}})$ labeled by $l=1,...,{M} ;\sum_{\bf j}$
stands for multiple sum over $j_{1},j_{2},...,j_{M}$. We apply the
last equation of (\ref{N16}) in our analytical approach, and the
second last equation in our numerical analysis.

\subsection*{\em 2.3 Projection synthesis and teleportation} %2.3

We are interested in the optical truncation schemes based on the
generalized QSD, shown in figure 2, or its simplified versions
depicted in figure 3(a),(b). The input mode $\hat{a}_4$, being in
an arbitrary pure state, given by (\ref{N01}), is truncated to a
qudit state, given by (\ref{N02}), in mode $\hat{b}_1$. To achieve
the truncation desired, we assume that modes $\hat{a}_1$,
$\hat{a}_2$, and $\hat{a}_3$ are in the Fock states $|n_1\rangle$,
$|n_2\rangle$, and $|n_3\rangle$, respectively. Thus, the total
four-mode input state is
\begin{eqnarray}
|\Psi \rangle =|n_1\rangle _{1}|n_2\rangle _{2}|n_3\rangle
_{3}|\psi \rangle _{4}\equiv |n_1n_2n_3\psi \rangle \label{N17}
\end{eqnarray}
which is transformed into the output state $|\Phi \rangle$,
according to (\ref{N12}). Now, photon-counting of the output modes
$\hat{b}_2$, $\hat{b}_3$, and $\hat{b}_4$ is performed yielding
$N_{2}$, $N_{3}$ and $N_{4}$ photons, respectively. If the total
number of detected photons is equal to the sum of photons in modes
$\hat{a}_1,\hat{a}_2,\hat{a}_3$, i.e.,
$N_{2}+N_{3}+N_{4}=n_1+n_2+n_3\equiv d-1$, then the total four-mode
output state $|\Phi \rangle $ is reduced to the following
single-mode state:
\begin{eqnarray}
|\phi\rangle &\equiv& |\phi^{N_{2}N_{3}N_{4}}_{n_1n_2n_3} \rangle
={\cal N} \,_{2}\langle N_{2}|\,_{3}\langle N_{3}|\,_{4}\langle
N_{4}|\Phi \rangle
\nonumber\\
&=& {\cal N} \sum_{n=0}^{d-1}\langle n N_{2}N_{3}N_{4}|\Phi \rangle
|n \rangle = {\cal N} \sum_{n=0}^{d-1} c^{(d)}_n
\gamma_{n}|n\rangle \label{N18}
\end{eqnarray}
with the amplitudes $c^{(d)}_n$ defined as
\begin{equation}
c^{(d)}_{n}({\bf T},{\mbf \xi}) = \langle n
N_{2}N_{3}N_{4}|\hat{U}|n_1n_2n_3n\rangle \label{N19}
\end{equation}
depending, in particular, on the beam splitter transmittances
${\bf T}\equiv [t_1^2, t_2^2,t_3^2, t_4^2, t_5^2]$ and phase
shifts ${\mbf \xi}\equiv [\xi_1, \xi_2, \xi_3, \xi_4, \xi_5]$ and
$\xi_6$. In the following, we present solutions for $c_n^{(d)}$
assuming $\xi_6=0$. Nevertheless, since the action of the phase
shifter $P_6$ with $\xi_6$ simply corresponds to the
transformation of $c_n^{(d)}$ into $c_n^{(d)} \exp(n\xi_6)$, one
can readily obtain the general solutions for any $\xi_6$.

Perfect truncation is achieved independently of the form of the
input state $|\psi\rangle$ if the amplitudes $c^{(d)}_{n}$ are
equal to each other for all $n \le d-1$ or, equivalently,
\begin{equation}
\Delta({\bf T},{\mbf \xi})\equiv \sum_{n=1}^{d-1} |c^{(d)}_n({\bf
T},{\mbf \xi})- c^{(d)}_0({\bf T},{\mbf \xi})|=0 \label{N20}
\end{equation}
for some properly chosen values of the BS transmittances ${\bf T}$
and phase shifts ${\mbf \xi}$. So, the problem is to find such
${\bf T}$ and ${\mbf \xi}$, for which the amplitudes $c^{(d)}_n$
satisfy condition (\ref{N20}). It is worth noting that, for a
possible experimental realization of the scheme, it is essential to
have BSs with variable transmittance. A possible solution is to
replace each of the BSs by a Mach-Zehnder interferometer composed
of two symmetric 50:50 BSs, two mirrors and two PSs
\cite{Reck94,Reck96,Paris00}.

Our numerical minimalization of $\Delta$ reveals that perfect
truncation using the generalized QSD can be realized for various
values of the BS transmittances and phase shifts, which however,
correspond to different probabilities of successful truncation.
Here, we focus on very simple analytical solutions rather than
numerically optimized approximate ones. In the following, we will
analyze in detail truncations up to six-dimensional qudits.

The described truncation process via the GQSD can be considered a
kind of the form-limited quantum teleportation of the first $d$
terms of the Fock-state expansion of the incident light in analogy
to the qutrit-limited teleportation via the Pegg-Phillips-Barnett
QSD discussed in
\cite{Pegg98,Villas99,Koniorczyk00,Miranowicz02,Babichev03}. Even a
brief analysis of figure 2, shows that no light from the input port
4 can reach the output port 1. In fact, this transformation is
based on the same principles of the quantum entanglement and the
Bell-state measurement (the projection postulate) as the original
Bennett {\em et al} teleportation scheme \cite{Bennett93}. The
multi-photon entangled state is created by the beam splitters and,
as required, the original state $|\psi\rangle$ is destroyed by the
Bell-state measurement implemented by the BSs 3-5 and detectors
2-4. Thus, by assuming that an incident light is already prepared
in a $d$-dimensional (up to $d$=6) qudit state $|\psi^{(d)}\rangle$
and the conditional measurement is successfully performed, then the
state is teleported from mode $\hat{a}_4$ to
$|\phi^{(d)}\rangle=|\psi^{(d)}\rangle$ in mode $\hat{b}_1$.
%---------------------------------------------------------------------------
\begin{figure}% scheme 3
\hspace*{0mm}(a)\hspace*{4cm}(b)

\vspace*{-1mm}
\centerline{\epsfxsize=5cm\epsfbox{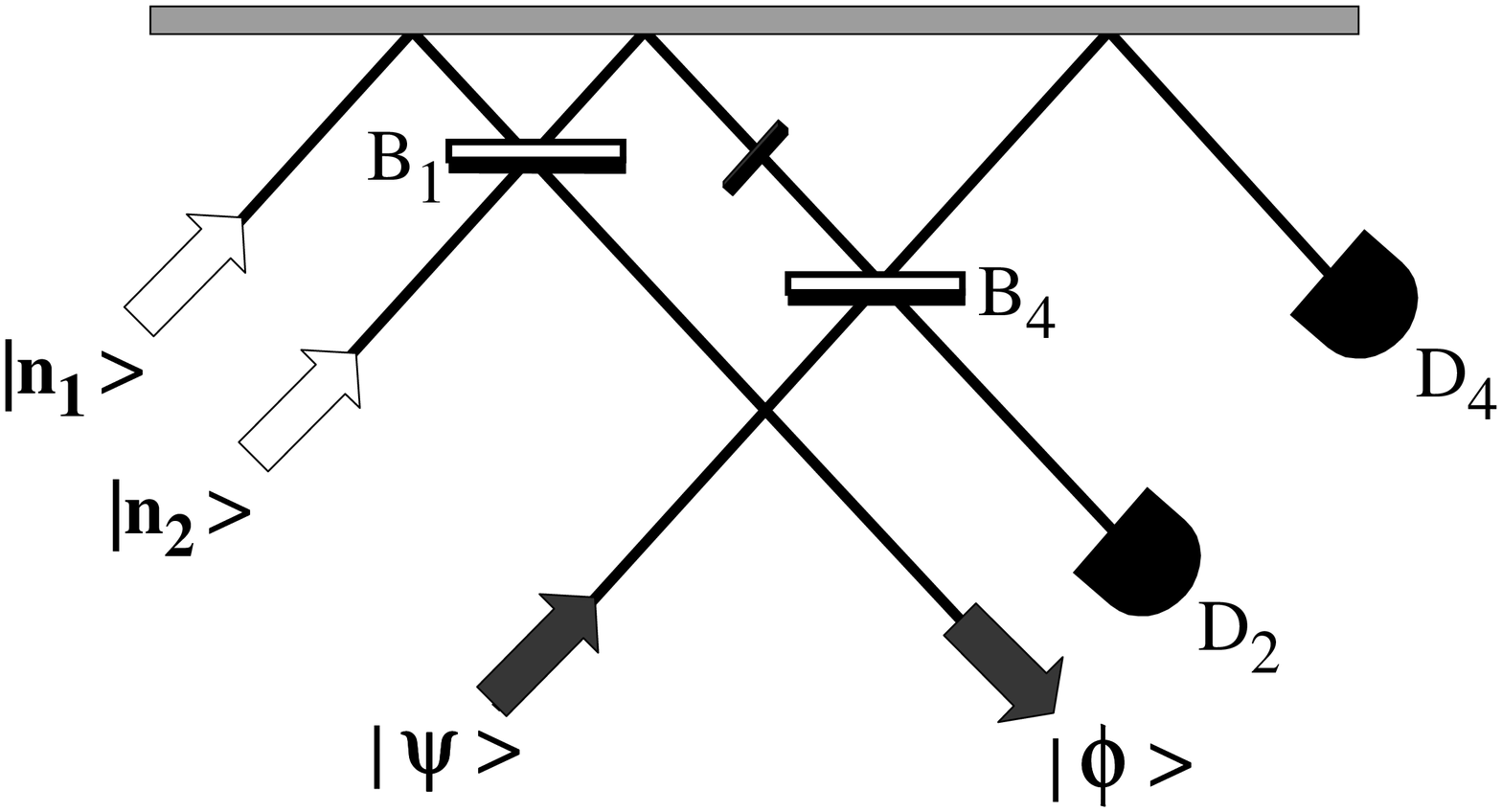}\hspace*{-7mm}
\epsfxsize=5cm\epsfbox{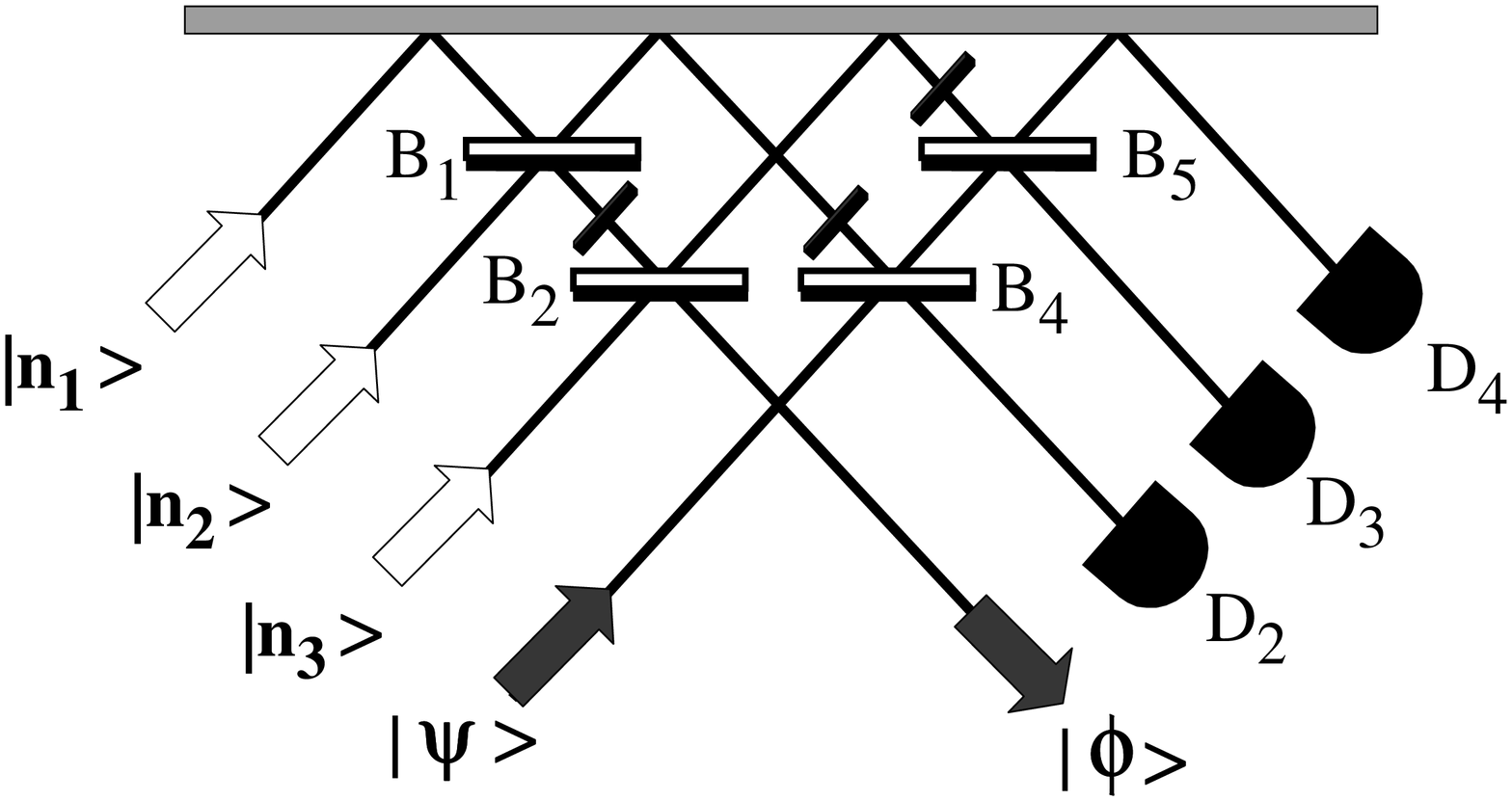}} \caption{Special cases of the
GQSD, shown in figure 2, corresponding to (a) the
Pegg-Phillips-Barnett QSD in figure 1, and (b) the scheme analyzed
in sections 4 and 6.}
\end{figure}

\section{Reductions to the Pegg-Phillips-Barnett QSD} %3

First, we show how the eight-port QSD can truncate the input state
$|\psi\rangle$, given by (\ref{N01}), to qubit state, given by
(\ref{N03}), as expected by the original Pegg-Phillips-Barnett
scissors device \cite{Pegg98,Barnett99}. The system shown in figure
1 is a special case of that shown in figure 2 by assuming, e.g.,
that BSs 1, 3 and 5 are perfectly reflecting, thus a complete set
of transmittances is ${\bf T}=[0,t_{2}^{2},0,t_{4}^{2},0]$. Note
that, in this configuration, the input port 1 and the output port 4
are unimportant, so can be neglected. Another way to generate the
truncated state (\ref{N03}) is to remove BSs 2, 3 and 5 as shown in
figure 3(a). Thus, the simplified version of the GQSD is described
by the transmittances
\begin{eqnarray}
{\bf T}=[t_{1}^{2},1,1,t_{4}^{2},1], \label{N21}
\end{eqnarray}
and we can also set that the only nonzero phase shift is $\xi_4$.
In this configuration, the input mode $\hat{a}_3$ is only
reflected from the large mirror and leaves the system without
interfering with the other modes. Thus, the truncation occurs
independently of the input state $|n_3\rangle$ and the photon
counting result in detector D3. Let $|\phi
_{n_{1}n_{2}}^{N_{2}N_{4}}\rangle$ denote the output state
$|\phi\rangle$ in mode $\hat{b}_1$ for the input states
$|n_{1}\rangle$, $|n_{2}\rangle$, $|\psi\rangle$ in modes
$\hat{a}_1$, $\hat{a}_2$, $\hat{a}_4$, together with detection of
$N_{2}$ and $N_{4}$ photons in detectors D2 and D4, respectively.
Note that equivalently one can analyze ${\bf T}=[1,t_{2}^{2},
1,1,t_{5}^{2}]$ instead of (\ref{N21}) to reduce the system to the
Pegg-Phillips-Barnett QSD. By applying (\ref{N07}) to (\ref{N16})
with transmittances given by (\ref{N21}), one readily finds the
output states:
\begin{eqnarray}
|\phi _{10}^{01}\rangle  &\sim&e^{i\xi _{4}}r_{1}r_{4}\gamma
_{0}|0\rangle
+t_{1}t_{4}\gamma _{1}|1\rangle,   \nonumber \\
|\phi _{01}^{10}\rangle  &\sim&e^{i\xi _{4}}t_{1}t_{4}\gamma
_{0}|0\rangle
+r_{1}r_{4}\gamma _{1}|1\rangle,   \nonumber \\
|\phi _{01}^{01}\rangle  &\sim&-e^{i\xi _{4}}t_{1}r_{4}\gamma
_{0}|0\rangle
+r_{1}t_{4}\gamma _{1}|1\rangle,  \label{N22}  \\
|\phi _{10}^{10}\rangle  &\sim&-e^{i\xi _{4}}r_{1}t_{4}\gamma
_{0}|0\rangle +t_{1}r_{4}\gamma _{1}|1\rangle.  \nonumber
\end{eqnarray}
As follows from (\ref{N22}), the states $|\phi _{10}^{01}\rangle$
and $|\phi _{01}^{10}\rangle$ become perfectly truncated qubit
states if $t_1=r_4$ and $\xi _{4}=0$. On the other hand, $|\phi
_{01}^{01}\rangle$ and $|\phi _{10}^{10}\rangle$ become (\ref{N03})
for the same transmittances of BSs 1 and 4 and phase shift $\xi
_{4}=\pi$. The optimized solution, giving the highest probability
of successful truncation, is found for the 50:50 BSs
($t_1^2=t_4^2=1/2$) in all 4 cases.  It is worth noting that in
\cite{Pegg98,Barnett99}, the internal phases of both BSs are chosen
as $\theta_{t}=0$ and $\theta_{r}=\pi/2$, so to obtain the exact
equivalence of the original scheme with ours, it is enough to set
the phase shift $\xi_4=\pi$.

As shown by Koniorczyk {\em et al} \cite{Koniorczyk00}, the
Pegg-Phillips-Barnett scissors device enables also the truncation
of an arbitrary incident state to the qutrit state
\begin{eqnarray}
|\phi^{(3)}_{\rm trun} \rangle  &\sim &\gamma _{0}|0\rangle +\gamma
_{1}|1\rangle +\gamma _{2}|2\rangle \label{N23}
\end{eqnarray}
which can be obtained in our setup by assuming the single-photon
Fock states in the input modes $\hat{a}_1$ and $\hat{a}_2$,
together with the single-photon counts in detectors D2 and D4.
Under these assumptions and by denoting $f'_i\equiv r_i^2-t_i^2$,
the output state in mode $\hat{b}_1$ becomes
\begin{eqnarray}
|\phi _{11}^{11}\rangle  \sim 2r_{1}t_{1}r_{4}t_{4}(e^{2i\xi
_{4}}\gamma _{0}|0\rangle +\gamma _{2}|2\rangle )+e^{i\xi
_{4}}f_{1}'f_{4}'\gamma _{1}|1\rangle \label{N24}
\end{eqnarray}
which is the desired truncated state if the parameters of BS1 and
BS4 are related as follows:
\begin{eqnarray}
t_{4}^{2}=\frac{1}{2}\left(1\pm
\frac{r_{1}t_{1}}{\sqrt{1-3(r_{1}t_{1})^{2}}}\right) \label{N25}
\end{eqnarray}
and the phase shift $\xi _{4}$ is equal to $\pi$ or zero,
respectively. By inspection of (\ref{N25}) one readily finds that
the optimum solutions occur for $t_{1}^{2}$ equal either to $(3-
\sqrt{3})/6\approx 0.21$ or to $(3+ \sqrt{3})/6\approx 0.79$ and
$t_{4}^{2}=t_{1}^{2}$ if $\xi_4=0$, in agreement with the results
of \cite{Koniorczyk00}, but also for $t_{4}^{2}=1-t_{1}^{2}$ if
$\xi_4=\pi$.

\section{Truncation to quartit states} %4

In this section, we demonstrate how to realize truncation of an
input state $|\psi\rangle$, given by (\ref{N01}), to the
four-dimensional qudit
\begin{eqnarray}
|\phi^{(4)}_{\rm trun} \rangle \sim \gamma_{0}|0\rangle +
\gamma_{1}|1\rangle +\gamma_{2}|2\rangle+\gamma_{3}|3\rangle
\label{N26}
\end{eqnarray}
referred to as the {\em quartit}. We apply the GQSD shown in figure
2, assuming that modes $\hat{a}_1$, $\hat{a}_2$, and $\hat{a}_3$
are the in single-photon Fock states, and single photons have been
measured in all detectors, $N_{2}=N_{3}=N_{4}=1$. If the light to
be truncated enters the interferometer in mode $\hat{a}_4$, then
the output state $|\phi \rangle=|\phi^{111}_{111} \rangle$ obtained
via the projection synthesis is given by (\ref{N17}) for $d=4$ and
the amplitudes
\begin{equation}
c^{(4)}_{n}\equiv \langle n111|\hat{U}|111n\rangle \label{N27}
\end{equation}
dependent on the BS and PS parameters. By applying the procedure
described in section 2, we find that the simplest solution is for
$n=d-1$ and reads as
\begin{eqnarray}
c^{(4)}_3&=&6 e^{2 i \xi_2} r_1 t_1 r_2 t_2^2 r_4 t_4^2 r_5 t_5.
\label{N28}
\end{eqnarray}
It is seen that $c^{(4)}_3$ is independent of the BS3 parameters,
so for simplicity let us assume that BS3 is removed
($t_3=1,\xi_3=0$). Thus, instead of a general setup, shown in
figure 2, we first analyze its simplified version shown in figure
3(b). Under the assumption, the solutions for the other amplitudes
are found to be
\begin{eqnarray}
c^{(4)}_0 &=& -2 e^{i (\xi_4 + \xi_5)} t_2 t_4 ( e^{i (\xi_2 +
\xi_5)} f'_1 r_2 r_5 t_5 + e^{i \xi_4} f'_5 r_1 t_1 r_4),\nonumber \\
c^{(4)}_1&=& -2 r_1 t_1 r_2 r_4 r_5 t_5 (e^{2 i (\xi_2 + \xi_5)}
f''_2 +
e^{2 i \xi_4} f''_4) \nonumber  \\
&&+e^{i (\xi_2 + \xi_4 + \xi_5)} f'_1 f'_2 f'_4 f'_5,\label{N29}
\\ c^{(4)}_2 &=& 2 e^{i \xi_2} t_2 t_4 (e^{i (\xi_2 + \xi_5)}
r_1 t_1 g_2 r_4 f'_5 + e^{i \xi_4} f'_1 r_2 g_4 r_5 t_5)\nonumber
\end{eqnarray}
where, for brevity, the $n$-primed $f$ denotes
$r_{k}^{2}-nt_{k}^{2}$, $g_k\equiv 2r_{k}^{2}-t_{k}^{2}$, and the
global phase factor $\exp(i\xi_1)$, the same for all $c^{(4)}_n$,
was omitted as it cancels out during the renormalization with
${\cal N}$. Our multiport interferometer can act as a good quantum
scissors device if there exist parameters ${\bf T}$ and ${\mbf
\xi}$ such that condition (\ref{N20}) is satisfied. As explained in
subsection 2.3, we focus on the simplicity of the solutions,
although we realize that they are not optimal as implied by the
results of our numerical experiments. For example, a simple
solution is found for the transmittances equal to
\begin{equation}
{\bf T}
=\left[\frac{1}{3},\frac{1}{4},1,\frac{1}{3},\frac{1}{2}\right]
\label{N30}
\end{equation}
and zero phase shifts ${\mbf \xi}$ except $\xi_5=\pi/2$. By
applying these values to (\ref{N29}), one readily finds that
$c^{(4)}_n=1/12$ for $n=0,1,2,3$. Another solution of (\ref{N20})
is found for the transmittances
\begin{equation}
{\bf T}=\left[\frac{1}{26}(13 - 3 \sqrt{13}), \frac{1}{2}, 1,
\frac{1}{3},\frac{1}{4}(2 + \sqrt{3})\right] \label{N31}
\end{equation}
and ${\mbf \xi}=0$, which results in a constant amplitude
$c^{(4)}_n=1/(4\sqrt{39})$, although much lower than that found
for the first solution. Numerically it is easy to find solutions,
which correspond to higher $c^{(4)}_n$ than those given by our
analytical solutions. E.g., by choosing
$T=[0.78494,0.69001,1,0.87451,0.70185]$ and zero phase shifts
${\mbf \xi}$ except $\xi_5=\pi$, then the amplitudes $c^{(4)}_n$
for the output state $|\phi \rangle=|\phi^{111}_{111} \rangle$ are
constant at the value of  0.134.

A closer look at the amplitudes (\ref{N29}) reveals that they
remain unchanged by some permutations of transmittances accompanied
by a proper change in the phase shifts, or by replacement of
$t^2_k$ by $1-t^2_k$. In particular, we find the following
relations:
\begin{eqnarray}
c^{(4)}_n({\bf T},{\mbf \xi})&=&
c^{(4)}_n([t_5^2,t_4^2,1,t_2^2,t_1^2],{\mbf \xi})
 \label{N32}
\end{eqnarray}
if $\xi_1=0,\xi_4=\xi_2+\xi_5$, and
\begin{eqnarray}
c^{(4)}_n({\bf T},{\mbf \xi})&=&
c^{(4)}_n([1-t_1^2,t_2^2,1,t_4^2,t_5^2],{\mbf
\xi'})\nonumber\\
&=&  c^{(4)}_n([t_1^2,t_2^2,1,t_4^2,1-t_5^2]{\mbf \xi''})
 \label{N33}
\end{eqnarray}
if ${\mbf \xi}={\mbf \xi'}={\mbf \xi''}$ except for
$\xi'_2=\xi_2+\pi$ or, equivalently, $\xi'_4=\xi_4+\pi$, and
$\xi''_5=\xi_5+\pi$. Thus, by transforming solutions (\ref{N30})
and (\ref{N31}) according to (\ref{N32}) and (\ref{N33}) one can
find new solutions.

\section{Selective truncations} %5

Here, we focus on generalized truncations of the input state
$|\psi\rangle$ into a finite superposition of the form
$|\phi^{(d)}_{\rm trun} \rangle$ albeit with some states (say
$|k_1\rangle$, $|k_2\rangle$, ...) removed, i.e.,
\begin{eqnarray}
|\psi\rangle \rightarrow |\phi^{(d)}_{{\rm holes}\,k_1,k_2,...}
\rangle ={\cal N} \sum_{n=0 \atop (n\ne k_1,k_2,...)
}^{d-1}\gamma_{n}|n\rangle \label{N34}
\end{eqnarray}
corresponding to the case when the amplitudes $c^{(d)}_n$ vanish
for $n=k_1,k_2,...$ and are constant but nonzero for the other
$n$. This kind of quantum state engineering can be interpreted as
the truncation with hole burning. In general, the {\em hole
burning} in the Fock space of a given state of light, according to
Baseia {\em et al} (see \cite{Escher04} and references therein)
and Gerry and Benmoussa \cite{Gerry02}, means selective removal of
one or more specific Fock states from the field. Although,
originally, the hole burning was applied to infinite-dimensional
states, given by (\ref{N01}), this concept can also be used in the
case of finite-dimensional states $|\phi^{(d)}_{\rm trun}
\rangle$. Alternatively, following the interpretation of D'Ariano
{\em et al} \cite{Dariano00}, one can refer to the above quantum
state engineering, especially when the number of holes exceeds
$(d-1)/2$, as a kind of {\em Fock-state filtering}, which enables
selection of some Fock states (say $|j_1\rangle$, $|j_2\rangle$,
...) from a given input state $|\psi\rangle$, i.e.,
\begin{eqnarray}
|\psi\rangle \rightarrow |\phi_{{\rm filter}\,j_1,j_2,...} \rangle
={\cal N} (\gamma_{j_1}|j_1\rangle+\gamma_{j_2}|j_2\rangle
+\cdots). \label{N35}
\end{eqnarray}
Here, we show how the GQSD can be used for the Fock-state
filtering and the hole burning in the case of $d=4$. As the first
example, we analyze the truncation to $|\phi^{(4)}_{\rm trun}
\rangle$ with the two-photon Fock state removed which results in
\begin{eqnarray}
|\phi^{(4)}_{{\rm hole}\,2} \rangle \sim \gamma_{0}|0\rangle
+\gamma_{1}|1\rangle+\gamma_{3}|3\rangle. \label{N36}
\end{eqnarray}
We find that state $|\phi^{111}_{111} \rangle$, with amplitudes
given by (\ref{N29}), is reduced to (\ref{N36}) for various
transmittances ${\bf T}$ and phase shifts ${\mbf \xi}$, including
the following:
\begin{equation}
{\bf T} =\Big[\frac{1}{14}(7+\sqrt{21}),\frac{1}{3},1,\frac{1}{2},
\frac{1}{10}(5-\sqrt{5})\Big] \label{N37}
\end{equation}
and ${\mbf \xi=0}$. Similarly, the other truncated states with a
single hole:
\begin{subequations} \label{N38}
\begin{eqnarray}
 |\phi^{(4)}_{{\rm hole}\, 0} \rangle &\sim& \gamma_{1}|1\rangle
+\gamma_{2}|2\rangle+\gamma_{3}|3\rangle,  \\
 |\phi^{(4)}_{{\rm hole}\, 1} \rangle &\sim& \gamma_{0}|0\rangle
+\gamma_{2}|2\rangle+\gamma_{3}|3\rangle,
\end{eqnarray}
\end{subequations}
can be generated by the GQSD from $|\phi^{111}_{111} \rangle$ if,
e.g., ${\mbf \xi=0}$ and the transmittances are as follows:
\begin{subequations} \label{N39}
\begin{eqnarray}
{\bf T}
&=&\Big[\frac{1}{14}(7+\sqrt{21}),\frac{1}{3},1,\frac{1}{2},
\frac{1}{4}(2-\sqrt{2}) \Big],
 \\
{\bf T} &=&\Big[\frac{1}{2}-\frac{3}{2}\sqrt{\frac{5}{173}}
,\frac{1}{2},1,\frac{1}{6},\frac{1}{2}+\frac{5}{2}\sqrt{\frac{3}{203}}
\Big],
\end{eqnarray}
\end{subequations}
respectively. One can check that superpositions of any two Fock
states $|k\rangle$ and $|l\rangle$ for $k,l=0,...,3$, i.e.,
\begin{eqnarray}
 |\phi_{{\rm filter}\, kl} \rangle &\sim& \gamma_{k}|k\rangle
+\gamma_{l}|l\rangle, \label{N40}
\end{eqnarray}
can be obtained as special cases of $|\phi^{111}_{111} \rangle$,
e.g., for ${\mbf \xi=0}$ and the transmittances given by
\begin{eqnarray}
|\phi_{{\rm filter}\, 02} \rangle:&& {\bf T}
=\Big[1,\frac{1}{2},1, 1,\frac{1}{2}\Big],
\nonumber \\
|\phi_{{\rm filter}\, 03} \rangle:&& {\bf T}
=\Big[\frac{1}{2}\Big(1-\sqrt{\frac{5}{133}}\Big),\frac{1}{2},1,
\frac{1}{6},\frac{1}{2}+\frac{3}{2}\sqrt{\frac{3}{155}}\Big],
\nonumber\\
|\phi_{{\rm filter}\, 13} \rangle:&& {\bf T}
=\Big[\frac{1}{2},\frac{1}{3}(3-\sqrt{3}),1,
\frac{1}{3}(3-\sqrt{3}),\frac{1}{2}\Big],
\label{N41}\\
|\phi_{{\rm filter}\, 23} \rangle:&& {\bf T}
=\Big[\frac{1}{2}\Big(1-\sqrt{\frac{5}{37}}\Big),\frac{1}{2},1,\frac{1}{6},
\frac{1}{2}\Big(1+\sqrt{\frac{3}{35}}\Big)\Big]. \nonumber
\end{eqnarray}
Note that our exemplary state $|\phi_{{\rm filter}\, 02} \rangle$,
given in (\ref{N41}), can be realized in the Pegg-Phillips-Barnett
scheme, since the chosen transmittances are a special case of
(\ref{N21}). All the above examples were found for $t_3=1$. But we
have not found solutions $|\phi_{{\rm filter}\, 12} \rangle$ by
assuming $t_3=1$, together with the single-photon states in the
input modes $\hat{a}_{1},\hat{a}_2,\hat{a}_3$ and the
single-photon counts in detectors D2, D3, and D4. While keeping
the latter two requirements, one can change only the transmittance
of BS3. Then, we find a solution
\begin{eqnarray}
|\phi_{{\rm filter}\, 12} \rangle:&& {\bf T}
=\Big[\frac{1}{2}+\frac{1}{\sqrt{5}},\frac{8}{9},\frac{1}{2},
\frac{1}{2}+\frac{1}{\sqrt{5}},1 \Big] \label{N42}
\end{eqnarray}
and ${\mbf \xi}=0$. On the other hand, the state $|\phi_{{\rm
filter}\, 12}\rangle$ can be realized in the QSD with $t_3=1$,
e.g., as a special case of the output state
$|\phi^{101}_{110}\rangle$ for
\begin{eqnarray}
|\phi_{{\rm filter}\, 12} \rangle:&& {\bf T}
=\Big[\frac{1}{10}(5-\sqrt{15}),\frac{2}{3},1,\frac{1}{2},
\frac{1}{2} \Big] \label{N43}
\end{eqnarray}
and ${\mbf \xi}=0$. The scheme enables also a synthesis of Fock
states via teleportation. From $|\phi^{111}_{111}\rangle$, one can
synthesize the two and three photon Fock states for the following
transmittances:
\begin{subequations} \label{N44}
\begin{eqnarray}
|2\rangle:&& {\bf T}
=\Big[1,\frac{1}{2},\frac{1}{3},\frac{1}{2},1\Big],
\\
|3\rangle:&& {\bf T}
=\Big[\frac{1}{2},\frac{1}{2},1,\frac{1}{2},\frac{1}{2}\Big]
\end{eqnarray}
\end{subequations}
and, e.g., ${\mbf \xi}=0$ except $\xi_5=\pi/2$ in the latter case.
Actually, with the choice (\ref{N44}a), the state $|2\rangle$ is
generated for arbitrary phase shifts. The two-photon Fock state
cannot be obtained from $|\phi^{111}_{111}\rangle$ assuming
$t_3=1$, which can be shown analytically. However, the state
$|2\rangle$ can easily be obtained even for $t_3=1$ but from other
states, e.g. $|\phi^{101}_{110}\rangle$. It is worth noting that
by applying the transformations given by (\ref{N32}) and
(\ref{N33}) to the above solutions for ${\bf T}$, one can easily
obtain new analytical solutions for the generation of states
$|\phi^{(4)}_{{\rm hole}\,k} \rangle$ and $|\phi_{{\rm
filter}\,kl} \rangle$. We have given only some specific examples
of {\bf T}, which guarantee the desired truncation. Although, it
is out of the main goal of this paper, it is possible to give more
general conditions for ${\bf T}$, e.g.: (i) for any ${\bf
T}=[t_1^2,t_2^2,1,1,1/2]$ with $t^2_1\ne 1/2$ and $t_2\ne 0,1$,
the output state is $|\phi_{{\rm filter}\, 02} \rangle$, (ii) for
any ${\bf T}=[1/2,t_2^2,1,1-1/(3r_2^2),1/2]$ with $t_2\in(0,2/3)$
the output state is $|\phi_{{\rm filter}\, 13} \rangle$, assuming
in both cases ${\mbf \xi}=0$. As already emphasized, the solutions
presented here are usually not optimized, but simple enough to
show analytically that the specific truncations can be realized by
the GQSD.

\section{Truncation to five-dimensional qudit states} %6

A question arises whether the GQSD, shown in figure 3(b) with the
removed BS3, can be used for truncation of the input state up to
more than quartits. So, first we analyze possibilities of the
truncation of an input state $|\psi \rangle$, given by (\ref{N01}),
to the qudit state $|\phi^{(5)}_{\rm trun}\rangle$ being a special
case of (\ref{N02}) for $d=5$.  As usual, we assume that light to
be truncated is in mode $\hat{a}_4$, and the input modes
$\hat{a}_1$ and $\hat{a}_3$ are in the single-photon states, but,
by contrast to the former sections, we choose mode $\hat{a}_2$ to
be in the two-photon state. So, the total input state is
$|\Psi\rangle=|121\psi\rangle$. We assume that the conditional
measurement yields the single-photon counts in detectors D2 and D4,
but the two-photon count in D3, thus the resulting output state
$|\phi \rangle=|\phi^{121}_{121}\rangle$ is given by (\ref{N18})
for $d=5$ and the amplitudes
\begin{equation}
c^{(5)}_{n}\equiv \langle n121|\hat{U}|121n\rangle \label{N45}
\end{equation}
equal to
\begin{eqnarray*}
c^{(5)}_{0} &=&e^{i(\xi _{4}+\xi _{5})}t_{2}t_{4}\big( 3e^{2i
 (\xi_{2}+\xi_{5})}f_{1}''r_{1}r_{2}^{2}r_{5}t_{5}^{2}   \\
&&+2e^{i(\xi _{2}+\xi _{4}+\xi _{5})}g_{1}
t_{1}r_{2}r_{4}g_{5}t_{5}+3e^{2i\xi
_{4}}r_{1}t_{1}^{2}r_{4}^{2}r_{5}f_{5}''\big),
\end{eqnarray*}
\begin{eqnarray*}
c^{(5)}_{1} &=&  3r_{1}t_{1}r_{2}r_{4}r_{5}t_{5}(e^{3i(\xi _{2}+\xi
_{5})}r_{1}f_{2}'''r_{2}t_{5}+e^{3i\xi
_{4}}t_{1}f_{4}'''r_{4}r_{5}) \\
&&-e^{i(\xi _{2}+\xi _{4}+\xi _{5})}(e^{i(\xi _{2}+\xi
_{5})}f_{1}''r_{1}f_{2}''r_{2}f_{4}'g_{5}t_{5} \\
 &&+e^{i\xi _{4}}g_{1}t_{1}f_{2}'f_{4}''r_{4}f_{5}''r_{5}) ,
\end{eqnarray*}
\begin{eqnarray*}
c^{(5)}_{2} &=& e^{i\xi _{2}}t_{2}t_{4}\{2t_{1}r_{2}r_{4}t_{5}[
3e^{2i\xi
_{4}}(r_{1}^{2}t_{4}^{2}-r_{1}^{2}g_{4}+t_{1}^{2}f_{4}')r_{5}^{2}
\\
&& -e^{2i(\xi _{2}+\xi _{5})}
r_{1}^{2}(f_{2}''g_{5}+g_{2}f_{5}'+g_{2}r_{5}^{2})]
\\
&&+e^{i(\xi _{2}+\xi _{4}+\xi _{5})}r_{1}r_{5}
[2r_{1}^{2}g_{2}r_{4}^{2}f_{5}'-2t_{1}^{2}
(f_{2}'+r_{2}^{2})g_{4}f_{5}''
\\
&&-r_{1}^{2}g_{2} (t_{4}^{2}f_{5}''+2r_{4}^{2}t_{5}^{2})]\},
\end{eqnarray*}
\begin{eqnarray*}
c^{(5)}_{3}&=&3e^{2i\xi
_{2}}r_{1}t_{2}^{2}t_{4}^{2}r_{5}\big[e^{i(\xi _{2}+\xi
_{5})}r_{1}t_{1}(3r_{2}^{2}-t_{2}^{2})r_{4}f_{5}''
\\
&&+e^{i\xi _{4}}\!
f_{1}''r_{2}(3r_{4}^{2}-t_{4}^{2})r_{5}t_{5}\big],
\end{eqnarray*}
\begin{eqnarray}
c^{(5)}_{4}=12e^{3i\xi
_{2}}r_{1}^{2}t_{1}r_{2}t_{2}^{3}r_{4}t_{4}^{3}t_{5}r_{5}^{2}.
\label{N46}
\end{eqnarray}
As in (\ref{N29}), the irrelevant global phase factor
$\exp(i\xi_1)$ was canceled out from all $c^{(5)}_n$ in
(\ref{N46}). We have not found analytical solutions for the BS and
PS parameters satisfying condition (\ref{N20}) with $d=5$ for the
amplitudes given by (\ref{N46}), as the problem requires finding
roots of 6th order equations. Thus, we have applied numerical
procedure for finding the BS parameters for which $\Delta = 0 $
with precision of the order of $10^{-16}$. We have found various
solutions including that for the transmittances equal to ${\bf
T}$=[0.30464, 0.38775, 1, 0.81740, 0.18438], $\xi_4=\pi$ and the
other phase shifts set to zero, which results in the constant
amplitude $c^{(5)}_n$ for $n=0,...,4$. By placing BS3 in the setup
with $t_3\neq 1$ one can find the other solutions with larger
constant $c^{(5)}_n$. However, since we are not interested in the
optimalization but rather simplicity of the scheme, we do not
present these solutions here.

\section{Truncation to six-dimensional qudit states} %7

The eight-port QSD, shown in figure 2, enables truncation of the
incident light $|\psi \rangle$ even to the six-dimensional qudit
state $|\phi^{(6)}_{\rm trun} \rangle$ as a special case of
(\ref{N02}) for $d=6$. To achieve the perfect truncation, we assume
a single-photon Fock state in mode $\hat{a}_2$, two-photon states
in modes $\hat{a}_1,\hat{a}_3$, and that the conditional
measurement yields $N_2=N_4=2$ and $N_3=1$ photon counts. Then, the
output state $|\phi \rangle=|\phi^{212}_{212}\rangle$ is given by
(\ref{N18}) for $d=6$ and the amplitudes
\begin{equation}
c^{(6)}_{n}\equiv \langle n212|\hat{U}|212n\rangle. \label{N47}
\end{equation}
The simplest-form amplitude is for $n=d-1$, which reads as
\begin{eqnarray}
c^{(6)}_5= 30 e^{2 i \xi_1+ 3 i \xi_2} r_1 t_1^2 r_2^2 t_2^3 r_4^2
t_4^3 r_5 t_5^2. \label{N48}
\end{eqnarray}
As in the former cases for the truncation to qudits with $d<6$, the
amplitude $c^{(d)}_{d-1}$ is independent of the BS3 parameters.
Unfortunately, contrary to the former cases, we have not found
numerically any solutions satisfying $c^{(6)}_n=$const for
$n=0,...,d-1$ for the simplified scheme with removed BS3, shown in
figure 3(b). Thus, we analyze again the general scheme shown in
figure 2 with $t_3\neq 0,1$. As an example, we give a solution for
$c^{(6)}_4$ to be as follows:
\begin{eqnarray}
c^{(6)}_4 &=& 6 e^{2 i (\xi_1+\xi_2)}  t_1 r_2 t_2^2 r_4 t_4^2 t_5
\{ (3 r_4^2 - 2 t_4^2)
 \nonumber \\
&\times& e^{i \xi_4} [e^{i \xi_2} r_1 t_1 (3 r_2^2 -2 t_2^2) r_3 +
e^{i \xi_3} g_1 r_2 t_3] r_5 t_5
 \nonumber \\ &+&
e^{i \xi_5} [e^{i \xi_2} r_1 t_1 (3 r_2^2 -2 t_2^2) t_3 -e^{i
\xi_3} g_1 r_2 r_3 ] r_4 g_5 \}.\qquad \label{N49}
\end{eqnarray}
Solutions for $c^{(6)}_n$ with $n=0,...,3$ are quite lengthy so,
we do not present them explicitly here. Nevertheless, they have
been used in our numerical search of the BS and PS parameters
satisfying condition (\ref{N20}) for $d=6$. Thus, we have found
various solutions for which $\Delta\sim 10^{-16}$. We just mention
a solution for the BS transmittances equal to ${\bf T}$ =[0.75572,
0.41783, 0.32503, 0.83274, 0.50338], $\xi_4=\pi$ and the other
phase shifts equal to zero, which results in the constant nonzero
amplitude $c^{(6)}_n$ for $n=0,...,5$, and obviously vanishing for
$n>5$. Another solution is found for the same phase shifts but
transmittances equal to ${\bf T}$=[0.58154, 0.28519, 0.46753,
0.68558, 0.49836], which results also in a constant $c^{(6)}_n$
but one that slightly lower than that in the former case.

\section{Imperfect photon counting} %8

Now, we address an important problem from an experimental point of
view that concerns the deterioration effects of system
imperfections on the fidelity of truncation and teleportation. To
analyze such effects one can follow the approaches of, for
example,
\cite{Barnett98,Barnett99,Villas99,Ozdemir01,Ozdemir02,Miranowicz04}
applied to the Pegg-Phillips-Barnett QSD. Here, we focus on
imperfect photodetection.

Photon counting in mode $\hat b_{i}$ by an imperfect detector with
a finite efficiency $\eta_i$ and a mean dark count rate $\nu_i$
can be described by a positive-operator-valued measure (POVM)
\cite{Peres93} with the following elements \cite{Barnett98}
%----------------------------------------------------------------------
\begin{equation}
\hat\Pi^{(b_{i})}_{N_i}= \sum_{n=0}^{N_i} \sum_{m=n}^\infty
\frac{e^{-\nu_i}\nu_i^{N_i-n}}{(N_i-n)!}
\eta_i^{n}(1-\eta_i)^{m-n}C^{m}_{n} |m\rangle_i\, _i\langle m|
\label{N50}
\end{equation}
summing up to the identity operator $\hat I$. In (\ref{N50}),
$N_i$ is the number of registered photocounts in detector $D_i$,
$n$ is the actual number of photons entering detector, and $N_i-n$
is the number of dark counts, $C^{m}_{n}$ are binomial
coefficients. The mean dark count rate $\nu$ in (\ref{N50}) is
related to the standard dark count rate $R_{\rm dark}$ by the
relation $\nu=\tau_{\rm res} R_{\rm dark}$, where  $\tau_{\rm
res}$ is the detector resolution time. In analogy to the analysis
of the Pegg-Phillips-Barnett QSD given in \cite{Ozdemir01}, we
compare the fidelities for the states truncated by the generalized
scissors in relation to three types of applied detectors:  (i)
Conventional photodetectors (e.g., avalanche photo-diodes, APDs)
providing only a binary answer to the question whether any photons
have been registered or not, thus described by the POVM with the
two elements:
%----------------------------------------------------------------------
\begin{equation}
\hat\Pi_{c0}^{(b_{i})} = \hat\Pi_0^{(b_{i})}, \quad
\hat\Pi_{c1}^{(b_{i})} = \hat I-\hat\Pi_0^{(b_{i})}.
 \label{N51}
\end{equation}
(ii) Single-photon resolving photodetectors (e.g., visual light
photon counters, VLPCs \cite{Takeuchi99}) providing a trinary
answer to the question whether zero, one or more than one photons
have been registered, so given by the POVM with the following
three elements:
%----------------------------------------------------------------------
\begin{equation}
\hat\Pi_{s0}^{(b_{i})} = \hat\Pi_0^{(b_{i})}, \quad
\hat\Pi_{s1}^{(b_{i})} = \hat\Pi_1^{(b_{i})}, \quad
\hat\Pi_{s2}^{(b_{i})} = \hat I
-\hat\Pi_0^{(b_{i})}-\hat\Pi_1^{(b_{i})}
 \label{N52}
\end{equation}
where $\hat\Pi_{0,1}^{(b_{i})}$ in (\ref{N51}) and (\ref{N52}) are
given by (\ref{N50}). (iii) And unrealistic detectors (labeled by
$r$) resolving any number of simultaneously absorbed photons
described by the POVM elements $\hat\Pi_{rN_i}^{(b_{i})}\equiv
\hat\Pi_{N_i}^{(b_{i})}$, given by (\ref{N50}) for any $N_i$. Such
detectors are not available, although some methods (including the
so-called photon chopping \cite{Paul96}) have been proposed to
measure photon statistics with conventional devices. By including
the imperfect photon counting by detectors D2, D3 and D4, the
output state at mode ${\hat b}_{1}$ can be described by the
following density matrix
%----------------------------------------------------------------------
\begin{eqnarray}
\hat{\rho}_{x}={\cal N}\; {\rm Tr}_{(b_{2},b_{3},b_{4})}
\left(\hat\Pi_{xN_2}^{(b_{2})} \hat\Pi_{xN_3}^{(b_{3})}
\hat\Pi_{xN_4}^{(b_{4})} |\Phi\rangle\langle\Phi|\right),
\label{N53}
\end{eqnarray}
where the partial trace is taken over the detected modes $\hat
b_{2}$, $\hat b_{3}$, and $\hat b_{4}$; $\hat\Pi_{xN_i}^{(b_{i})}$
are the POVM elements for a given type of detectors $x=c,s,r$;
$|\Phi\rangle$ is the four-mode output state, given by
(\ref{N12}), and ${\cal N}$ is the normalization. For simplicity,
we can assume identical detectors with $\eta\equiv
\eta_2=\eta_3=\eta_4$ and $\nu\equiv \nu_2=\nu_3=\nu_4$. Deviation
of the realistically truncated state $\hat{\rho}_{x}$ from the
ideally truncated state $|\phi \rangle$ is usually described by
the fidelity
%----------------------------------------------------------------------
\begin{equation}
F_x=\langle \phi| \hat{\rho}_{x} | \phi \rangle. \label{N54}
\end{equation}
In our numerical analysis we assume: (i) $\eta=0.7$ and $R_{\rm
dark}\sim 100\,{\rm s}^{-1}$ for conventional detectors (see e.g.
\cite{Ozdemir01}), (ii) $\eta=0.88$ and $R_{\rm dark}\sim
10^4\,{\rm s}^{-1}$ for single-photon detectors (VLPCs) as
experimentally achieved by Takeuchi {\em et al} \cite{Takeuchi99},
(iii) for theoretic photon-number resolving detectors we choose
the same $\eta$ and $R_{\rm dark}$ as in (ii). Moreover, we put
$\tau_{\rm res}\sim 10$ ns. We observe that the truncation
fidelity in the system with imperfect photodetection depends on
the chosen transmittances. In particular, different solutions
described in section 4 for perfect truncation (with $F=1$) in the
lossless system correspond not only to different probabilities of
success but also to different fidelities of the truncation in the
lossy system. For $\alpha=0.4$, we find that the fidelity for the
truncation up to quartit states in the system described by the
transmittances given below equation (\ref{N31}) drop from one to
$F_c\approx 0.91$ for the conventional detectors and to
$F_s=F_r\approx 0.98$ for the VLPCs and the photon-number
resolving detectors. The fidelities of the truncation up to
five-dimensional states in the system described in section 6 are
estimated to be $F_c\approx 0.67$ for the conventional detectors,
$F_s\approx 0.95$ for the VLPCs, and $F_r\approx 0.96$ for the
photon-number resolving detectors if $\alpha=0.4$. These
estimations show that the conventional photodetectors can
effectively be used for the low-intensity-field truncations
described in sections 3--5, where at most single-photon detections
are required. In the schemes described in sections 6 and 7, where
detections of two-photons are important, one has to apply at least
single-photon resolving detectors even in the low-intensity limit.
It is worth noting that by increasing the number of detectors and
beam splitters in the discussed pyramid configuration, one can
achieve truncations to higher-dimensional states by detecting no
or single photons only \cite{Miranowicz05}. In our estimation we
have assumed, based on \cite{Takeuchi99}, relatively high values
of the dark count rates. However, it has recently been
experimentally demonstrated by Babichev {\em et al}, by rigorously
synchronizing the photon count events, that the dark counts can be
reduced to a negligible level \cite{Babichev03}. As multiport
optical interferometers have already been experimentally realized
\cite{Mattle95,Reck96,Mohseni04}, thus it seems that the proposed
GQSD for the truncation and teleportation of at least quartit
states is accessible to experiments with present-day technology
for low-intensity incident fields.

\section{Discussion and conclusions} %9

We are aware that our analysis of the quantum state truncation via
a GQSD is not yet complete . Among open problems to be analyzed to
a greater detail we should mention: (i) A generalization of the
scheme for the truncation of an arbitrary $d$-dimensional qudit
states based on the $2N$-port interferometer in the triangle
configuration with the top BS removed. By symmetry of the scheme,
shown in figure 2, such generalization is straightforward, e.g.,
along the lines of \cite{Reck94}. However, it would be desirable
to find the minimum number of detections in the multiport QSD,
which enables the state truncation to a qudit state of a given
dimensionality. (ii) An analysis of other kinds of losses
(including mode mismatch in addition to imperfect photon-counting)
in the generalized optical state truncation. (iii) A detailed
experimental proposal of the scheme for the truncation at least to
qutrit and quartit states. (iv) A hard problem is the
optimalization of solutions for the BS parameters to obtain the
highest probability of the truncation for $d\ge 4$. These problems
are being under our current investigation \cite{Miranowicz05}.

In conclusion, we have proposed a generalization of the
Pegg-Phillips-Barnett six-port QSD (shown in figure 1) to the
eight-port optical interferometer, depicted in figure 2. The
analyzed system enables, upon post-selection based on photon
counting results, generation and teleportation of qudit states
(for $d=2,...,6$) by truncation of an input optical field at the
$(d-1)$th term of its Fock-state expansion. We have discussed
examples of selective truncations, which enable Fock-state
filtering and hole burning in the Fock space of an input optical
field. We have also analyzed deterioration of the truncation
fidelity due to realistic photon counting including finite
photon-number resolution, inefficiency and dark counts of
photodetectors. Our estimations suggest that the scheme is
experimentally feasible at least for the generation and
teleportation of quartit states of low-intensity incident fields.

{\bf Acknowledgments}. The author thanks Ji\v{r}\'\i{} Bajer,
Steve Barnett, Andrzej Grudka, Nobuyuki Imoto, Masato Koashi,
Wies\l{}aw Leo\'nski, Yu-xi Liu, \c{S}ahin \"Ozdemir and Ryszard
Tana\'s for sound collaboration on implementations of quantum
scissors. This work was supported by the Polish State Committee
for Scientific Research under grant No. 1 P03B 064 28.

\end{document}